\documentstyle[11pt,epsfig]{article}
\newcommand{\bea}{\begin{eqnarray}}
\newcommand{\eea}{\end{eqnarray}}

\def\simlt{\stackrel{<}{{}_\sim}}
\def\simgt{\stackrel{>}{{}_\sim}}

\begin{document}
\begin{titlepage}

\thispagestyle{empty}

\vspace{0.2cm}

\title{ The CP properties of the lightest Higgs boson  
 with sbottom effects
\author{M\"{u}ge Boz\\
Hacettepe University, Department of Physics,\\  
06532 Ankara, Turkey\\}}
\date{}
\maketitle

\begin{center}\begin{minipage}{5in}

\begin{center} ABSTRACT\end{center}
\baselineskip 0.2in
{In the framework of the recently proposed gluino-axion model,
using the effective potential method and taking into account
the top-stop as well as the bottom-sbottom effects,
we discuss  the CP--properties of the lightest Higgs boson,
in particular its CP--odd composition,
which can offer new opportunities at  collider searches. 
It is found that although the CP-odd composition of the lightest Higgs
increases slightly with  the inclusion of the sbottom effects,
it  never exceeds $\%0.17$ for all values of the renormalization scale  Q
ranging from top mass to ${\rm  TeV}$ scale}
\end{minipage}
\end{center}
\end{titlepage}

\eject
\rm
\baselineskip=0.25in

\section{Introduction}
The radiative corrections to the masses of Higgs bosons in the MSSM have
received much attention from the beginning, and an important step in the
understanding of the MSSM Higgs sector was the significant modification of the tree
level bound  by the radiative corrections, dominated by the top quark and
top squark loops  \cite{haber,okada,ellis1,ellis2}. The radiative corrections have
been computed by using different approximations such as the effective
potential \cite{okada,ellis1,ellis2}, and  diagrammatic
\cite{haber,hempfling} methods. More complete treatment  
of these results include the
complete one-loop on-shell renormalization
\cite{chandowski}, the renormalization group (RG) improvement for resumming the
leading logarithms \cite{okada2,sasaki,haber2},
the iteration of the RG  equations to two--loops
with the use of the effective potential techniques\cite{casas1,
carena2,carena3,haber3}
and  two loop on-shell renormalization  \cite{espinosa,carena4,heinmeyer}.

In the recent literature, the studies on the radiatively induced CP
violation effects and the theoretical predictions for the Higgs boson masses
and couplings as functions of relevant minimal SUSY model(MSSM) parameters has been
carried out  in several  directions.
For example, in \cite{pilaftsis1,pilaftsis4}, the implications of  the presence of CP 
phases in the soft
SUSY breaking sector allowing to the mixing of CP even and CP odd states
were discussed.
More recently  the  mass matrix of the neutral Higgs bosons
of the MSSM has been calculated using the effective potential method,
in the case of the small splittings between
squark mass eigenstates in  \cite{pilaftsis2},
taking into account only the dominant   top-stop contributions in
 \cite{demir2} and including the sbottom contributions in   \cite{ham}.
Additional contributions from the chargino, W and  the charged Higgs exchange loops
were computed in \cite{ibrahim}. In \cite{choi1}, one-loop corrections to the mass matrix
of the neutral Higgs bosons in the
MSSM were calculated by using the effective potential method for an
arbitrary splitting between squark masses,  including the electroweak and
gauge couplings and the leading two loop corrections.
More complete treatment of the effective Higgs potential in the MSSM
including the two-loop leading logarithms induced by top-bottom
Yukawa couplings as well as those associated with QCD corrections by
means of RG methods were performed in \cite{carena}.

It is a well-known fact that 
in the supersymmetric (SUSY) extensions of S.M, apart from the physical
phases $\delta_{CKM}$ and $\theta_{QCD}$, also existing in the SM,
there appear novel sources of CP violation via the phases of the soft
supersymmetry breaking mass terms \cite{dugan}. Besides  their contribution to the known 
CP-violating observables
such as  electric dipole moments of the particles \cite{edm1, edm2}, these phases  
also induce CP
violation in the  Higgs sector \cite{pilaftsis1,pilaftsis4,
pilaftsis2,demir2,demir1,demir7}.
Moreover, the SUSY theories which are designed to solve the hierarchy problems,
possess two hierarchy problems:  One concerning the strong CP problem, also
existing in the S.M, whose source is the neutron EDM exceeding the present
bounds by nine orders of magnitude  \cite{harris} and the other is the $\mu$ puzzle,
concerning the Higgsino Dirac mass parameter $\mu$ which follows from the
superpotential of the model. A   simultaneous solution to these two
hierarchy problems can be achieved by the gluino-axion
model\cite{demir3,demir4,ma} with a new kind
of axion \cite{peccei,kim,shifman,dine}, which couples to the gluino rather than to quarks.
Besides,  the low energy theory is identical to
the MSSM with all sources of the soft SUSY phases.
Due to all these abilities of the model, in the analysis below we shall
adopt its parameter space.

In the recent literature,
the radiative corrections to the Higgs masses and mixings,
dominated by the top-stop contributions have been studied in gluino-axion model \cite{boz}.
In Ref. \cite{boz}, which was based on Ref. \cite{demir2},
the mass squared matrix of Higgs scalars
involves the  one loop function $g(x,y)$, as well as the other
scale-independent terms. The function  $g(x,y)$ 
is related to $f(x,y)$ in such a way that  $g(x,y)=f(x,y)-log\frac{xy}{Q^4}$,
where $f(x,y)$ is a scale-dependent loop function
whose expression is given by $f(x,y)=-2+log\frac{xy}{Q^4}+\frac{y+x}{y-x}log\frac{y}{x}$.
However, since the explicit dependence of Q in the
one-loop function $f(x,y)$ is actually cancelled  by the explicit  Q dependence of the function
$g(x,y)$, then, $g(x,y)=-2+\frac{y+x}{y-x}log\frac{y}{x}$ does not have an explicit dependence
on the renormalization scale Q, unlike $f(x,y)$.
Therefore, in both of the works of \cite{demir2,boz}
the elements of  the mass squared matrix of Higgs
scalars, which depend only on  the scale-independent function $g(x,y)$ as well as
the other scale-independent terms, do not have an explicit dependence on Q.
Actually, in \cite{demir2,boz}
the one-loop bottom-sbottom contributions to $\Delta V$ are not included,
and the terms proportional to $\mu$, $A_t$
are obtained by neglecting the D-terms in the stop masses,
to gain independence of the renormalization scale Q,
since the D-term contributions to the squark
masses are quite small.
On the other hand,  in Ref. \cite{ham}
the one-loop bottom sbottom contributions as well as the
top-stop contributions to radiative corrections are taken into consideration.
Thus, the  elements of the mass squared mass matrix of the Higgs scalars
not only include the scale-independent function
$g(x,y)$ with various parameters, but have   
an explicit dependence on the renormalization scale
through the scale dependent function
$f(x,y)$, and the scale dependent logarithmic terms, 
stemming from the additional 
contributions to the radiative corrections. 

The main purpose of this work is  to analyze the CP violation effects
on the lightest Higgs boson in the framework of the gluino-axion model
using the recent experimental data  \cite{LEP},
by the inclusion of the bottom-sbottom effects, as well as the top- stop contributions.
As the  Q-dependence is taken into consideration,  
we particularly adress the issues,  whether or not the  various 
renormalization scales, all being around the weak scale,
would  lead us to a large amount of CP violation opportunities,
and, investigate whether one can find an 
appropriate limit of  reasonable agreement  with
the scale-independent results \cite{boz}.
We will base our calculations to those of Ref. \cite{ham},
However, we differ from \cite{ham} in the sense that 
in our analysis,  all the chosen parameters  are
specific to the gluino-axion model, namely
all  the soft mass parameters in this theory are fixed in terms of the $\mu$ parameter.

The organization of this work is as follows: In Sec.~2, starting from the
Higgs sector structure of the gluino-axion model, we compute the
$(3\times3)$ dimensional mass matrix of the Higgs scalars whose
elements  are expressed in terms of the parameters of  the model
under concern. In Sec.~3,  we make the numerical
analysis to study the CP violation effects on the lightest Higgs boson,
and  analyze the influence of Q on its  CP--odd composion, as well as 
on $\mu$. We conclude in Sec.~4.

\section{The model}
In the gluino-axion  model,
the invariance of the supersymmetric Lagrangian and  the
invariance of all supersymmetry breaking terms   under
$U(1)_{R}$  are guaranteed by promoting the ordinary $\mu$ operator to
a   composite operator containing the singlet
composite superfield  $\widehat{S}$ with unit R charge  \cite{demir3}.
When the scalar component of the singlet develops a
vacuum expectation value around the Peccei- Quinn scale $v_s\sim
10^{11}~{\rm GeV}$, an effective $\mu$-parameter $\sim \mbox{a TeV}$ is induced such that
\begin{eqnarray}
\label{mu}
\mu\equiv v_s^2/M_{Pl}\times e^{-i \theta_{QCD}/3}\sim \mbox{a TeV}\times e^{-i \theta_{QCD}/3}
\end{eqnarray}
where $\theta_{QCD}$ is
the effective QCD vacuum angle. Therefore, the vacuum expectation value of the singlet serves for two 
important purposes for the model under concern: Its magnitude determines the scale of supersymmetry 
breaking and
its phase solves the strong CP--problem.

The effective Lagrangian at low-energy is given by  \cite{demir3}:
\begin{eqnarray}
\label{softMSSM}
{\cal{L}}^{soft}_{MSSM}&=& \tilde{Q}^{\dagger}M_{Q}^{2} \tilde{Q} + 
\tilde{u^c}^{\dagger} M_{u^c}^{2} \tilde{u^c}+\tilde{d^c}^{\dagger}
M_{d^c}^{2}\tilde{d^c}+\tilde{L}^{\dagger} M_{L}^{2} \tilde{L}+
\tilde{e^c}^{\dagger} M_{e^c}^{2} \tilde{e^c}\nonumber\\&+&
\Big\{ A_{u} \tilde{Q}\cdot {H}_{u}~\tilde{u^c}+A_{d} \tilde{Q}\cdot 
{H}_{d} ~\tilde{d^c} + A_{e} \tilde{L}\cdot {H}_{d} ~\tilde{e^c}\big] + h. c.
\Big\}\nonumber \\ &+& M_{H_u}^{2} |H_u|^{2}+M_{H_d}^{2} |H_d|^{2}+
\left(\mu\ B H_{u}\cdot H_{d} + h. c. \right)\nonumber\\
&+&\Big\{M_{3} \tilde{\lambda}^{a}_{3}\tilde{\lambda}^{a}_{3} +
M_2 \tilde{\lambda}^{i}_{2}\tilde{\lambda}^{i}_{2}+ M_1 \tilde{\lambda}_{1}
\tilde{\lambda}_{1}+ h. c. \Big\},
\end{eqnarray}
The soft terms of the low energy Lagrangian in the gluino-axion model are identical to those in the 
general MSSM except for the fact that the soft masses are all expressed in terms of
the $\mu$ parameter through appropriate flavour matrices 
except for the fact that the soft masses are all expressed in terms of
the $\mu$ parameter through appropriate flavour matrices. 
The flavour 
matrices form the sources of CP violation and intergenerational mixings in 
the squark sector. The phases of the trilinear couplings ($A_{u,d,e}$), 
the gaugino masses ($M_{3,2,1}$), and the effective $\mu$--parameter defined
in  (\ref{mu}) are the only phases which can generate CP violation observables. 

It is known that the dominant contributions to the  one-loop radiative corrections
come from the the top quark and top squark loops as long as  
$\tan\beta\simlt 50$, as in the CP-conserving case
\cite{haber,okada,ellis1,ellis2}.
However, for a more sensitive calculation, 
we will  take into account of the contributions from the  bottom- sbottom,  
as well as the  top-stop  quark loops. 
For  convenience, we set 
the soft SUSY breaking scalar-quark
masses as $M_{\tilde{Q}} =M_{\tilde{u}}=M_{\tilde{d}}$,
and  the squark trilinear couplings as $A_{t}=A_{b}$. Then, the explicit
expressions  for the mass parameters  in (\ref{softMSSM})
defined  as follows: The top (bottom) squark soft masses are  given by:
\begin{eqnarray}
M_{\tilde{Q}}^{2}=k_{Q}^{2}\ |\mu|^{2}~,
\end{eqnarray}
where $k_{Q}$ is a real parameter.
The top(bottom) squark trilinear couplings read as:
\begin{eqnarray}
\label{at}
A_{t}=\mu \ k_t^{*},
\end{eqnarray}
where $k_t$ is a complex parameter. 
The  tree level Higgs soft masses are defined by:
\begin{eqnarray}
M_{H_u}^{2}=y_u |\mu^{2}|~,\ \ \ M_{H_d}^{2}=y_d |\mu ^{2}|~,\ \ \
\mu\ B=|\mu|^{2} (\frac{8 m_s^{2}}{v_{s}^{2}}+k_{\mu})~,
\end{eqnarray}
where 
$m_s^2\sim v_s^2$ is a natural choice as discussed in  \cite{demir3}.  Here $y_u$ and $y_d$ are 
real parameters, and $k_{\mu}$ is a complex parameter determining the phase of 
the $B$ parameter. 
which  can be identified with the relative phase of the Higgs doublets \cite{demir2}.

After electroweak breaking the Higgs doublets in (\ref{softMSSM}) can be expanded as 
\begin{eqnarray}
\label{doublet}
H_{d}&=&\left(\begin{array}{c c} H_{d}^{0}\\
H_{d}^{-}\end{array}\right)=\frac{1}{\sqrt{2}}\left(\begin{array}{c c}
v_{d}+\phi_{1}+i\varphi_{1}\\ H_{d}^{-}\end{array}\right)\;,\nonumber\\
H_{u}&=&\left(\begin{array}{c c} H_{u}^{+}\\
H_{u}^{0}\end{array}\right)=\frac{e^{i\theta}}{\sqrt{2}}\left(\begin{array}{c c}
H_{u}^{+}\\ v_{u}+\phi_{2}+i\varphi_{2}\end{array}\right)\; . 
\end{eqnarray}
where $\tan\beta\equiv v_u/v_d$ as usual, and the angle parameter $\theta$ is 
the misalignment between the two Higgs doublets. 

We follow the effective potential method for computing the one-loop
corrected Higgs masses and mixings. As usual, the  entries of the Higgs masses and their mixings 
can be calculated up to one loop accuracy by the second derivatives of the
effective potential with respect to the Higgs fields.
\begin{eqnarray}
M^{2}=\left(\frac{\partial^{2}\ V} {\partial \chi_{i} \partial \chi_{j}}\right)_{0}\,,
\mbox{where}\; 
\chi_{i} \in {\cal{B}}=\{\phi_{1}, \phi_{2}, \varphi_{1}, \varphi_{2}\} \; ,
\end{eqnarray}
where $V\equiv V_0 + V_{1-loop}$ is the radiatively corrected Higgs
potential \cite{ham}, and as we mentioned before we take into account the 
top-stop and bottom-sbottom loop corrections.
The stop and sbottom mass-squared eigenvalues are 
given by: 
\begin{eqnarray}
\label{ccc}
m_{\tilde{t}_{1,2}}^{2}&=&\frac{1}{4}M_{Z}^2\, c_{ 2 \beta}
+ 
m_{t}^2+ k_{Q}^{2}|\mu|^{2} \mp  \Delta_{\tilde{t}}^{2}~,
\end{eqnarray}
\begin{eqnarray}
m_{\tilde{b}_{1,2}}^{2}&=&-\frac{1}{4}M_{Z}^2\, c_{ 2 \beta}
+ 
m_{b}^2+ k_{Q}^{2}|\mu|^{2} \mp  \Delta_{\tilde{b}}^{2}~,
\end{eqnarray}
with,
\begin{eqnarray}
\label{deltat}
\Delta_{\tilde{t}}^{2}&=&\sqrt{ \Big( \frac{2}{3} M_{W}^2- \frac{5}{12}
M_{Z}^2\Big)^{2} c^{2}_{ 2
\beta}+
m_{t}^{2}|\mu|^2 \Big(|k_{t}|^{2}+t^{-2}_{\beta}-2 |k_{t}| t^{-1}_{\beta} \,c_{
\varphi_{kt}}\Big)}~,
\end{eqnarray}
\begin{eqnarray}
\label{delb}
\Delta_{\tilde{b}}^{2}&=&\sqrt{ \Big( \frac{1}{12} M_{Z}^2- \frac{1}{3}
M_{W}^2\Big)^{2} c^{2}_{ 2
\beta}+
m_{b}^{2} |\mu|^2 \Big(|k_{t}|^{2}+t^ {2}_{\beta}-2 |k_{t}| t_{\beta} \,c_{
\varphi_{kt}}\Big)}~,
\end{eqnarray}
where 
$c_{\beta}$=$\cos\beta$,\,\,$c_{\varphi_{kt}}$=$\cos \varphi_{kt}$,
$s_{\beta}$=$\sin\beta$,\,\,$s_{\varphi_{kt}}$=$\sin \varphi_{kt}$,
$t_{\beta}$=$\tan\beta$,\,\, $t^{-1}_{\beta}$=$\cot\beta$.
The stop and sbottom mass splittings depend explicitely on the  total CP violation angle
$\varphi_{kt}$ such that
\begin{eqnarray}
\varphi_{kt}=\mbox{Arg}[\mu A_{t}^{*}]=\mbox{Arg}[k_t]~,
\end{eqnarray}
where $k_{t}$ has been defined in (\ref{at}). 

The  (3$\times$3) dimensional Higgs mass--squared matrix  can be expressed as: 
\begin{eqnarray}
\label{massmat}
\left(\begin{array}{c c c}
M_{Z}^{2} c^{2}_{\beta} + \tilde{M}_{A}^{2} s^{2}_{\beta}+\Delta_{11} & 
-(M_{Z}^{2}+\tilde{M}_{A}^{2})s_{\beta}c_{\beta}+\Delta _{12} & \Delta _{13}\\
-(M_{Z}^{2}+\tilde{M}_{A}^{2})s_{\beta}c_{\beta} +\Delta _{12} 
& M_{Z}^{2} s^{2}_{ \beta}  +\tilde{M}_{A}^{2} c^{2}_{\beta}
+\Delta _{22} &\Delta _{23}\\
\Delta _{13} & \Delta _{23} & \tilde{M}_{A}^{2}+\Delta _{33}\\
\end{array}\right)
\end{eqnarray}
in the basis ${\cal{B}}=\{\phi_{1}, \phi_{2},
\sin\beta \varphi_{1}+\cos\beta \varphi_{2}\}$  using (\ref{doublet}). 

The elements of the radiatively corrected mass--squared matrix read as: 

\begin{eqnarray}
\Delta_{11t}&=&
\frac{\beta_{k}}{2}
\Bigg[ 
 \frac{ M_{Z}^4 c^{2}_{\beta}}{8}
\log\frac{m_{\tilde{t}_{2}}^{2} m_{\tilde{t}_{1}}^{2}}{Q^4}
+
\frac{M_{Z}^2 c_{\beta}}{2}
\frac{( 2 m_{t}^2 \mu {\cal{ R}}_{t1}+ \frac{1}{2} s_{ 2\beta}  
{\cal{X}}_{t})} {  s_{ \beta}
(m_{\tilde{t}_{2}}^{2}-m_{\tilde{t}_{1}}^{2})}
\log\frac{m_{\tilde{t}_{2}}^{2}} {m_{\tilde{t}_{1}}^{2}}
\nonumber\\
&-&
\frac{1}{2}\Big( 
\frac{2 k_t m_{t}^2 \mu^{2} c_{\varphi_{kt}}}
{s_ { 2 \beta}} 
-{\cal{Z}}_{t}(M_W,\ M_Z) c^{2}_ {\beta} \Big)  
f(m_{\tilde{t}_{1}}^{2},m_{\tilde{t}_{2}}^2)
\nonumber\\
&-&
\frac{( 2 m_{t}^2 \mu {\cal{ R}}_{t1}+\frac{1}{2} s_{ 2 \beta} 
{\cal{X}}_{t})^2}{ 2 s^{2}_{\beta}
(m_{\tilde{t}_{2}}^{2}-m_{\tilde{t}_{1}}^{2})^2} 
g(m_{\tilde{t}_{1}}^{2},m_{\tilde{t}_{2}}^2)
\Bigg]
\end{eqnarray}

\begin{eqnarray}
\Delta_{11b}&=&
\frac{\beta_{k}}{2}\Bigg[ 
-\frac{4m_b^4}
{c^{2}_{\beta}}\log\frac{m_b^2}
{Q^2}+
\frac{1}{2}\left(\frac{2 m_b^{2}}{c_{ \beta}}-\frac{M_{Z}^2 c_{ \beta}}{2}\right)^2 
\log\frac{m_{\tilde{b}_{2}}^{2} m_{\tilde{b}_{1}}^{2}}{Q^4}
\nonumber\\
&+&
\frac{1}{2}\Big(\frac{4 m_b^{2}}{c_{  \beta}}-M_{Z}^2 c_{ \beta}\Big )
\frac{( 2 k_t m_{b}^2 \mu{\cal{ R}}_{b2}-   c^{2}_{ \beta} 
{\cal{X}}_{b})}{ c_{\beta} (m_{\tilde{b}_{2}}^{2}-m_{\tilde{b}_{1}}^{2})}
\log\frac{m_{\tilde{b}_{2}}^{2}}{ m_{\tilde{b}_{1}}^{2}}\nonumber\\
&-&
\frac{1}{2}\Big(\frac{ k_t m_{b}^2 \mu^{2} t_{ \beta} c_{\varphi_{kt}}}
{c^{ 2} _{\beta}}
-{\cal{Z}}_{b}(M_W,\ M_Z ) c^{2}_{ \beta}\Big) 
f(m_{\tilde{b}_{1}}^{2},m_{\tilde{b}_{2}}^2)\nonumber\\
&-&
\frac{( 2 k_t m_{b}^2 \mu {\cal{ R}}_{b2}-c^ { 2} _{\beta}
{\cal{X}}_{b})^2}{ 2 c^2_{ \beta}
(m_{\tilde{b}_{2}}^{2}-m_{\tilde{b}_{1}}^{2})^2}  
g(m_{\tilde{b}_{1}}^{2},m_{\tilde{b}_{2}}^2)\Bigg]~.
\end{eqnarray}

\begin{eqnarray}
\Delta_{22t}&=&\frac{\beta_{k}}{2} \Bigg[ 
-\frac{ 4  m_{t}^4}{ s^{2}_{\beta}} \log\frac{m_{t}^2}{Q^2}+
\frac{1}{2} 
\left (\frac{2 m_t^{2}}{s_{ \beta}}-\frac{M_{Z}^2 s_{ \beta}}{2}\right)^2 
\log\frac{m_{\tilde{t}_{2}}^{2} m_{\tilde{t}_{1}}^{2}}{Q^4}
\nonumber\\
&+&
\frac{1}{2}\Big(\frac{4 m_t^{2}}{s_{  \beta}}-M_{Z}^2 s_{\beta}\Big) 
\frac{( 2 k_t m_{t}^2 \mu {\cal{ R}}_{t2}-  s^{2}_{ \beta} 
{\cal{X}}_{t})}{  s_{ \beta} (m_{\tilde{t}_{2}}^{2}-m_{\tilde{t}_{1}}^{2})}
\log\frac{m_{\tilde{t}_{2}}^{2}} {m_{\tilde{t}_{1}}^{2}}
\nonumber\\
&-&
\frac{1}{2}\Big(   \frac{ k_t m_{t}^2 \mu^{2} t^{-1}_{\beta} c_{\varphi_{kt}}}
{s^ {2}_{ \beta}} 
-{\cal{Z}}_{t}(M_W,\ M_Z) s^{2}_{ \beta}\Big)  
f(m_{\tilde{t}_{1}}^{2},m_{\tilde{t}_{2}}^2)\nonumber\\
&-&
\frac{( 2 k_t m_{t}^2 \mu {\cal{R}}_{t2}-s^{2}_{ \beta}
{\cal{X}}_{t})^2}{ 2 s^{2}_{\beta}
(m_{\tilde{t}_{2}}^{2}-m_{\tilde{t}_{1}}^{2})^2} 
g(m_{\tilde{t}_{1}}^{2},m_{\tilde{t}_{2}}^2)\Bigg]~.
\end{eqnarray}

\begin{eqnarray}
\Delta_{22b}&=&\frac{\beta_{k}}{2}\Bigg[ 
\frac{ M_{Z}^4 s^{2}_{\beta}}{8} 
\log\frac{m_{\tilde{b}_{2}}^{2} m_{\tilde{b}_{1}}^{2}}{Q^4}
+
\frac{M_{Z}^2 c_{\beta}}{2} 
\frac{( 2 m_{b}^2 \mu {\cal{ R}}_{b1}+ \frac{1}{2}  s_{ 2  \beta} 
{\cal{X}}_{b})} {c_{\beta} (m_{\tilde{b}_{2}}^{2}-m_{\tilde{b}_{1}}^{2})} 
\log\frac{m_{\tilde{b}_{2}}^{2}} {m_{\tilde{b}_{1}}^{2}}\nonumber\\
&-&
\frac{1}{2}\Big(  \frac{2 k_t m_{b}^2 \mu^{2} c_{\varphi_{kt}}}
{s_ { 2 \beta}} 
-{\cal{Z}}_{b}(M_W,\ M_Z) s^{2}_{ \beta} \Big) 
f(m_{\tilde{b}_{1}}^{2},m_{\tilde{b}_{2}}^2)\nonumber\\
&-&
\frac{( 2 m_{b}^2 \mu {\cal{ R}}_{b1}+\frac{1}{2} s_{ 2 \beta} 
{\cal{X}}_{b})^2}{ 2 c^{2}_{\beta}
(m_{\tilde{b}_{2}}^{2}-m_{\tilde{b}_{1}}^{2})^2} 
g(m_{\tilde{b}_{1}}^{2},m_{\tilde{b}_{2}})
\Bigg]
\end{eqnarray}

\begin{eqnarray}
\Delta_{33t}&=&\frac{\beta_{k}}{2} \Bigg [- \frac{ 2 k_{t}^2 m_{t}^4 \mu^4 
s^{2}_{\varphi_{kt}}} 
{ s^{4}_{\beta} (m_{\tilde{t}_{2}}^{2}-m_{\tilde{t}_{1}}^{2})^2} 
g(m_{\tilde{t}_{1}}^{2},m_{\tilde{t}_{2}})
-\frac{1}{2}\frac{k_{t} m_{t}^2\mu^2  c_{ \varphi_{kt}}}  
{ s^{3}_{\beta} c_{\beta}}
f(m_{\tilde{t}_{1}}^{2},m_{\tilde{t}_{2}}^2) \Bigg]
\end{eqnarray}

\begin{eqnarray}
\Delta_{33b}&=&\frac{\beta_{k}}{2}\Bigg[ -\frac{2 k_{t}^2 m_{b}^4 \mu^4  
s^{2}_{\varphi_{kt}}} 
{ c^{4}_{\beta} (m_{\tilde{b}_{2}}^{2}-m_{\tilde{b}_{1}}^{2})^2} 
g(m_{\tilde{b}_{1}}^{2},m_{\tilde{b}_{2}})
-
\frac{1}{2}\frac{ k_{t} m_{b}^2\mu^2  c_ {\varphi_{kt}}}  
{ c^{3}_{\beta} s_{\beta}}
f(m_{\tilde{b}_{1}}^{2},m_{\tilde{b}_{2}}^2)\Bigg]
\end{eqnarray}

\begin{eqnarray}
\Delta_{12t}&=&\frac{\beta_{k}}{2}\Bigg[ 
\frac{ M_Z^2 s_{ 2 \beta}}{16} 
\Big(\frac{4 m_{t}^2}{s^2_{\beta}}-M_{Z}^2\Big)
\log\frac{m_{\tilde{t}_{2}}^{2} m_{\tilde{t}_{1}}^{2}}{Q^4}
\nonumber\\
&+&
\frac{ M_{Z}^2 c_{\beta}}{4} 
\frac{( 2 k_t m_{t}^2 \mu {\cal{ R}}_{t2}-  s^{2 }_{ \beta} 
{\cal{X}}_{t})} {  s_{\beta}
(m_{\tilde{t}_{2}}^{2}-m_{\tilde{t}_{1}}^{2})}
\log\frac{m_{\tilde{t}_{2}}^{2}}{ m_{\tilde{t}_{1}}^{2}}+
\nonumber\\
&+&
\frac{ 1}{4} 
\Big(\frac{4 m_{t}^2}{s_{\beta}}-M_{Z}^2 s_{\beta}\Big)
\frac{(2 m_{t}^2 \mu {\cal{ R}}_{t1}+ \frac{1}{2}  s_{ 2   \beta} 
{\cal{X}}_{t})} {   s_{\beta} (m_{\tilde{t}_{2}}^{2}-m_{\tilde{t}_{1}}^{2})} 
\log\frac{m_{\tilde{t}_{2}}^{2}}{ m_{\tilde{t}_{1}}^{2}}\nonumber\\
&+&
\frac{1}{4} \Big(  \frac{ 2 k_t m_{t}^2 \mu^{2}  c_{\varphi_{kt}}}
{s^{ 2}_{ \beta}}-\frac{1}{2}
{\cal{Z}}_{t}(M_W,\ M_Z)
s_{ 2 \beta} \Big) 
f(m_{\tilde{t}_{1}}^{2},m_{\tilde{t}_{2}}^2)\nonumber\\
&-&
\frac{( 2 m_{t}^2 \mu {\cal{ R}}_{t1}+ \frac{1}{2} s_{ 2 \beta}
{\cal{X}}_{t}) 
( 2 k_{t} m_{t}^2 \mu {\cal{ R}}_{t2}- s^ {2}_{ \beta}
{\cal{X}}_{t})}{ 2 s^{2}_{\beta}
(m_{\tilde{t}_{2}}^{2}-m_{\tilde{t}_{1}}^{2})^2}
g(m_{\tilde{t}_{1}}^{2},m_{\tilde{t}_{2}}^2)\Bigg]
\end{eqnarray}

\begin{eqnarray}
\Delta_{12b}&=&\frac{\beta_{k}}{2}\Bigg[ 
\frac{ M_Z^2 s _{ 2 \beta}}{16} 
\Big(\frac{4 m_{b}^2}{c^2_{\beta}}-M_{Z}^2\Big)
\log\frac{m_{\tilde{b}_{2}}^{2} m_{\tilde{b}_{1}}^{2}}{Q^4}\nonumber\\
&+&
\frac{ M_{Z}^2 s_{\beta}}{4} 
\frac{( 2  m_{b}^2 k_t \mu  {\cal{ R}}_{b2}-   c^{2}_{  \beta} 
{\cal{X}}_{b})} {  c_{ \beta} (m_{\tilde{b}_{2}}^{2}-m_{\tilde{b}_{1}}^{2})} 
\log \frac{m_{b_{2}}^2}{ m_{b_{1}}^2}\nonumber\\
&+&
\frac{ 1}{4} 
\Big(\frac{4 m_{b}^2}{c_{\beta}}-M_{Z}^2 c_{\beta}\Big)
\frac{( 2  m_{b}^2 \mu {\cal{ R}}_{b1}+   c^2_{ \beta} 
{\cal{X}}_{b})} {  c_{\beta} (m_{\tilde{b}_{2}}^{2}-m_{\tilde{b}_{1}}^{2})} 
\log \frac{m_{b_{2}}^2}{ m_{b_{1}}^2}\nonumber\\
&+&
\frac{1 }{4} \Big( \frac{ 2 k_t m_{b}^2 \mu^2  c_{ \varphi_{kt}}}
{c^ {2}_{ \beta}}
-{\cal{Z}}_{b}(M_W,\ M_Z)
s_{ 2 \beta} \Big) 
f(m_{\tilde{b}_{1}}^{2},m_{\tilde{b}_{2}}^2)\nonumber\\
&-&
\frac{( 2 m_{b}^2 \mu {\cal{ R}}_{b1}+ c^2_{ \beta}
{\cal{X}}_{b}) 
( 2 k_{t} m_{b}^2 \mu {\cal{ R}}_{b2}- c^2 _{\beta}
{\cal{X}}_{b})}{ 2 c^2_{ \beta}
(m_{\tilde{b}_{2}}^{2}-m_{\tilde{b}_{1}}^{2})^2}
g(m_{\tilde{b}_{1}}^{2},m_{\tilde{b}_{2}}^2)\Bigg]
\end{eqnarray}

\begin{eqnarray}
 \Delta_{13t}&=&\frac{\beta_{k}}{2}
\Bigg[
\frac{ k_{t} m_{t}^2 \mu^2 t^{-1}_{\beta} s_{ \varphi_{kt}}}{ s_{ \beta}} 
\Big(-\frac{1}{2}f(m_{\tilde{t}_{1}}^{2},m_{\tilde{t}_{2}}^2)-
\frac{ M_{Z}^2 \log \frac{m_{t_{2}}^2}{ m_{t_{1}}^2}}{2
(m_{\tilde{t}_{2}}^{2}-m_{\tilde{t}_{1}}^{2})}\Big)\nonumber\\
&+&
\frac{ k_{t} m_{t}^2 \mu^2  s_{ \varphi_{kt}} 
\left( m_{t}^2 \mu {\cal{ R}}_{t1}+ \frac{1}{2} s_{ 2 \beta}
{\cal{X}}_{t}\right) }{  s^{3}_{ \beta}
(m_{\tilde{t}_{2}}^{2}-m_{\tilde{t}_{1}}^{2})^2}
g(m_{\tilde{t}_{1}}^{2},m_{\tilde{t}_{2}})\Bigg]
\end{eqnarray}

\begin{eqnarray}
\Delta_{13b}&=&\frac{\beta_{k}}{2}\Bigg[
\frac{  k_{t} m_{b}^2 \mu^2 s_{ \varphi_{kt}}}{ c_{ \beta}} 
\Big(-\frac{1}{2}f(m_{\tilde{b}_{1}}^{2},m_{\tilde{b}_{2}}^2)-
( \frac{2 m_{b}^2} {c^{2}_{\beta}}- \frac{ M_{Z}^2}{2})
\frac{  \log \frac{m_{b_{2}}^2}{ m_{b_{1}}^2} }{
(m_{\tilde{b}_{2}}^{2}-m_{\tilde{b}_{1}}^{2})}\Big)\nonumber\\
&+&
\frac{  k_{t} m_{b}^2 \mu^2  s_{ \varphi_{kt}} 
\left( k_t m_{b}^2 \mu {\cal{ R}}_{b2}-c^{2}_{ \beta}
{\cal{X}}_{b}\right) }{  c^{3}_{ \beta} (m_{\tilde{b}_{2}}^{2}-m_{\tilde{b}_{1}}^{2})^2 }
g(m_{\tilde{b}_{1}}^{2},m_{\tilde{b}_{2}})\Bigg]
\end{eqnarray}

\begin{eqnarray}
\Delta_{23t}&=&\frac{\beta_{k}}{2}\Bigg[ 
\frac{  k_{t} m_{t}^2 \mu^2  s_{ \varphi_{kt}}}{  s_{ \beta}} 
\Big(-\frac{1}{2} f(m_{\tilde{t}_{1}}^{2},m_{\tilde{t}_{2}}^2)-
( \frac{2 m_{t}^2} {s^{2}_{\beta}}- \frac{ M_{Z}^2}{2})
\frac{  \log \frac{m_{t_{2}}^2}{ m_{t_{1}}^2} }{
(m_{\tilde{t}_{2}}^{2}-m_{\tilde{t}_{1}}^{2})}\Big)\nonumber\\
&+&
\frac{  k_{t} m_{t}^2 \mu^2  s_{ \varphi_{kt}} 
\left(2 k_t m_{t}^2 \mu {\cal{ R}}_{t2}-s^{2}_{ \beta}
{\cal{X}}_{t} \right)}{  s^3_{ \beta} (m_{\tilde{t}_{2}}^{2}-m_{\tilde{t}_{1}}^{2})^2}
g(m_{\tilde{t}_{1}}^{2},m_{\tilde{t}_{2}})\Bigg]
\end{eqnarray}

\begin{eqnarray}
\Delta_{23b}&=&\frac{\beta_{k}}{2}
\Bigg[\frac{  k_{t} m_{b}^2 \mu^2  t_{ \beta} s_ {\varphi_{kt}}}{c_{ \beta}} 
\Big(-\frac{1}{2}f(m_{\tilde{b}_{1}}^{2},m_{\tilde{b}_{2}}^2)-
\frac{ M_{Z}^2 \log \frac{m_{b_{2}}^2}{ m_{b_{1}}^2}}{2
(m_{\tilde{b}_{2}}^{2}-m_{\tilde{b}_{1}}^{2})}\Big)\nonumber\\
&+&
\frac{  k_{t} m_{b}^2 \mu^2  s_{ \varphi_{kt}} 
\left( m_{b}^2 \mu {\cal{ R}}_{b1}+ c^2_{ \beta}
{\cal{X}}_{b}\right)}{ c^3_{ \beta} (m_{\tilde{b}_{2}}^{2}-m_{\tilde{b}_{1}}^{2})^2}
g(m_{\tilde{b}_{1}}^{2},m_{\tilde{b}_{2}})\Bigg]
\end{eqnarray}
where $\beta_{k}=\frac {3}{ 8  \pi^2 v^2} $
and Q is the renormalization scale in the MS scheme. In the above expressions,
\begin{eqnarray}
{\cal{R}}_{t1}&=&\mu \left(|k_{t}|c_ {\varphi_{kt}} +
t^{-1}_{\beta}\right)~\,\,\,\,\,
{\cal{ R}}_{t2}=\mu\left(|k_{t}| + t^{-1}_{\beta} c_{\varphi_{kt}}\right)~,\nonumber\\
{\cal{R}}_{b1}&=&\mu\left(|k_{t}|c_{\varphi_{kt}} +
t_{\beta} \,\,\,\, \right)~\,\,\,\,\,
{\cal{R}}_{b2}=\mu\left(|k_{t}| + t_{\beta}\,  c_{\varphi_{kt}}\,\,\, \right)~.
\end{eqnarray}
and,
\begin{eqnarray}
{\cal{X}}_{t}&=&{\cal{Z}}_{t}(M_W,\ M_Z) c_{ 2 \beta} ~,\,\,\,\,\,
{\cal{X}}_{b}=- {\cal{Z}}_{b}(M_W,\ M_Z) c_{ 2 \beta} ~.
\end{eqnarray}
where
\begin{eqnarray}
{\cal{Z}}_{t}(M_W, M_Z)&=&(\frac{4}{3} M_{W}^2- \frac{5}{6} M_{Z}^2)^2~, 
\nonumber\\
{\cal{Z}}_{b}(M_W, M_Z)&=&(\frac{2}{3} M_{W}^2- \frac{1}{6} M_{Z}^2)^2~,
\end{eqnarray}
and the functions
$f(m_{\tilde{t}_{1}(\tilde{b}_{1})}^{2},m_{\tilde{t}_{2}(\tilde{b}_{2})}^2)$
and
$g(m_{\tilde{t}_{1}(\tilde{b}_{1})}^{2},m_{\tilde{t}_{2}(\tilde{b}_{2})}^2)$
are given by:

\begin{eqnarray}
f(m_{\tilde{t}_{1}(\tilde{b}_{1})}^{2},m_{\tilde{t}_{2}(\tilde{b}_{2})}^2)
&=&-2+\log\frac{
m_{\tilde{t}_{1}(\tilde{b}_{1})}^{2}m_{\tilde{t}_{2}(\tilde{b}_{2})}^2}{Q^4}+
\nonumber\\&&
\frac{
m_{\tilde{t}_{2}(\tilde{b}_{2})}^{2}+m_{\tilde{t}_{1}(\tilde{b}_{1})}^{2}}
{m_{\tilde{t}_{2}(\tilde{b}_{2})}^{2}-m_{\tilde{t}_{1}(\tilde{b}_{1})}^{2}}
\log\frac{{m_{\tilde{t}_{2}(\tilde{b}_{2})}^{2}}}
{{m_{\tilde{t}_{2}(\tilde{b}_{2})}^{2}}}
\end{eqnarray}

\begin{eqnarray}
g(m_{\tilde{t}_{1}(\tilde{b}_{1})}^{2},m_{\tilde{t}_{2}(\tilde{b}_{2})}^2)&=&
-2+\frac{m_{\tilde{t}_{2}(\tilde{b}_{2})}^{2}+m_{\tilde{t}_{1}(\tilde{b}_{1})}^2}
{m_{\tilde{t}_{2}(\tilde{b}_{2})}^{2}-m_{\tilde{t}_{1}(\tilde{b}_{1})}^2}
\log\frac{m_{\tilde{t}_{2}(\tilde{b}_{2})}^2}{m_{\tilde{t}_{1}(\tilde{b}_{1})}^2}
\end{eqnarray}

We diagonalize the Higgs mass--squared matrix (12)  by the similarity 
transformation
\begin{eqnarray}
{\cal{R}}M^{2}{\cal{R}}^{T}= {\rm diag}(m_{h_{1}}^{2},
m_{h_{2}}^{2}, m_{h_{3}}^{2})~,
\end{eqnarray}
where  ${\cal{R}}{\cal{R}}^{T}=1$, and  we define $h_3$ to be the lightest of all three.

One of the most important quantities in our analyses is 
the percentage CP composition of a given mass--eigenstate Higgs boson.
The percentage  CP compositions of the Higgs bosons in terms of the basis
elements are defined by 
\begin{eqnarray}
\rho_{i}=100\times |{\cal{R}}_{1i}|^{2};\,\,\, i=1, 2, 3.
\end{eqnarray}
where $\rho_{1}$, $\rho_{2}$  and $\rho_{3}$ correspond respectively
the $\phi_{1}$, $\phi_{2}$,
$\sin\beta \varphi_{1}+\cos\beta \varphi_{2}$  components of the Higgs boson
under concern. In the following, one of the main concerns will be the CP--odd
composition $\rho_{3}$, of the lightest Higgs boson as it can offer new
opportunities at colliders for observing the Higgs
boson\cite{demir7,choi2,choi6,choi7}.
We  will discuss  the  dependence of $\rho_3$
on  different  renormalization scales, as well as the interdependence of Q-$\mu$,
using the the previous theoretical \cite{boz}, as well as the experimental
results \cite{LEP}.

\section{Numerical Analysis}
In this section, we mainly adress  the phenomenological
consequences of explicit radiative CP violation effects in the
lightest Higgs boson, whose CP--odd composition as well as its mass are of
prime importance in direct Higgs boson searches at high-energy colliders.
As a result of  the standart model Higgs boson
searches at LEP, the lower bound on the lightest Higgs mass
is $115~{\rm GeV}$ (and correspondingly $\tan\beta \simgt 3.5 $) \cite{LEP}.
On the other hand, theoretically  the lightest Higgs boson mass can not exceed
$130~{\rm GeV}$ for large $\tan\beta$ \cite{heinmeyer}. Therefore, from the searches at LEP2,
the lower limit on mass of the SM Higgs boson excludes the
substantial part of the MSSM parameter space  particularly (for $m_t=175$
$\mbox{GeV}$), at small $\tan\beta$ ($\tan\beta \simlt 3.5$) \cite{LEP}. 

Being  a reflecting property of the underlying model, all the soft masses are
expressed in terms of the $\mu$ parameter, and since the $\mu$ parameter is already
stabilized to the weak scale  as a consequence of the naturalness, all dimensionless
quantities are expected to be ${\cal{O}}(1)$. Therefore,
we take  all the dimensionless quantities in the ${\cal{O}}(1)$.
In our numerical analysis, we study
two specific  values $\tan\beta$, namely  $\tan\beta=4$ and  $\tan\beta=30$,
to analyze the CP violation effects
on  the lightest Higgs boson in the the low and high $\tan\beta$ regimes.
\begin{figure}[htb]
\centering
\epsfxsize=6in
\epsffile{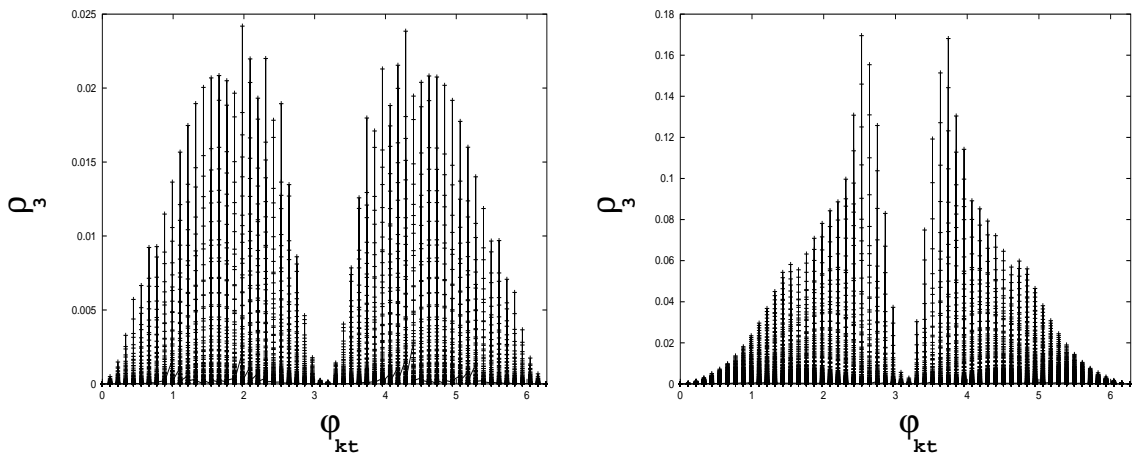}
\caption{The CP--odd composition ($\rho_{3}$) of the lightest Higgs boson as a 
function of $\varphi_{kt}$, 
for  $\tan\beta=4$ (left panel), and $\tan\beta=30$ (right panel),
where {\rm Q} varies from  $175 ~{\rm GeV}\, \mbox{to}\, 1200 ~{\rm GeV}$.}
\label{fig1}
\end{figure}

Shown in Fig.~1, is the  $\varphi_{kt}$ dependence of $\rho_{3}$,
designating the percentage CP--odd  compositions  of the lightest
Higgs boson ($h_{3}$), for $\tan\beta=4$ (left panel), and
$\tan\beta=30$ (right panel). In both panels Q changes
from $175 ~{\rm GeV} \,\, \mbox{to}\,\, 1200 ~{\rm GeV}$,
and  $|\mu|$  from $100~{\rm GeV}$ to $1000~\mbox{GeV}$
in the full $\varphi_{kt}$ range. Therefore, the  variation
of $\rho_3$ is represented by many points in the parameter space which  correspond to different values
of Q, and different values of $\mu$.  
As is seen from both panels of Fig.~1, the maximal value of $\rho_{3}$ occurs at  
$\%0.025$ for $\tan\beta=4$, and at $\%0.17$ for $\tan\beta=30$.
Moreover, a comparative look at both windows shows that,
although $\rho_{3}$ increases relatively with the increasing $\tan\beta$,
it never exceeds $\%0.17$ in the full  $\varphi_{kt}$ range,
for all values of Q changing from
from $175 ~{\rm GeV} \,\, \mbox{to}\,\, 1200 ~{\rm GeV}$.
Compared to its CP--even compositions, which form the 
remaining percentage, this CP--odd component is extremely small to cause
observable effects. It may, however, be still important when the radiative 
corrections to gauge and Higgs boson vertices are included  \cite{demir1}.

The analysis of Fig.~1, gives a general  idea of the
variation of $\rho_3$ in the full $\varphi_{kt}$ range.
However, since the parameter space is quite large, for a better understanding of
the properties of
the model under concern, we  focus on the different portions of parameter space
corresponding to the different values of Q in Fig.~2,
Therefore, in the left panel of Fig.~2,
we choose three specific values of
Q corresponding to  $ 325 ~{\rm GeV}$ \, $(''+'' )$, \, $600  ~{\rm GeV}$\,
$(''\times'')$, $1000 ~{\rm GeV}$ \, $(''.'')$, for $\tan\beta=4$,
and analyze the variation of  $\rho_3$ in the full $\varphi_{kt}$ range,
 for all values of $\mu$
changing from $100~{\rm GeV}$ to $1000~\mbox{GeV}$. On the other hand,
we  carry out the same analysis for
$\tan\beta=30$ in the right panel of Fig.~2. However,
for convenience, starting from a lower value of Q,
we let $190 ~{\rm GeV}$\, $(''\diamond'' )$, \,$ 325  ~{\rm GeV}$\,
$(''+'')$ ,\, $1000 ~{\rm GeV}$ \, $(''.'')$ values of Q  for this case. 
\begin{figure}[htb]
\centering
\epsfxsize=6in
\epsffile{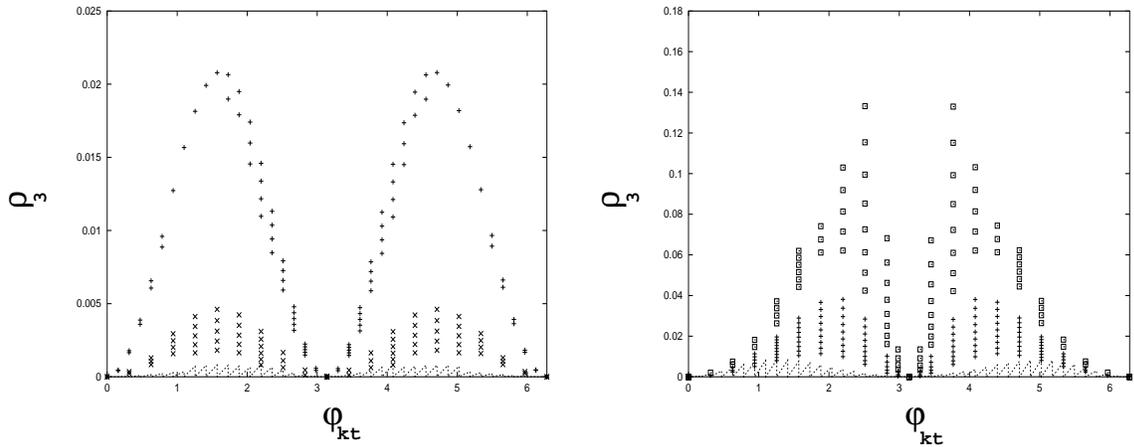}
\caption{The CP--odd composition ($\rho_{3}$) of the lightest Higgs boson as a
function of $\varphi_{kt}$, 
for  ${\rm Q}= 325 ~{\rm GeV} \, (''+'' ), \,  600~{\rm GeV}\,
(''\times''),  1000 ~{\rm GeV} \, (''.'')$,
for  $\tan\beta=4$ (left panel), and 
${\rm Q}=190 ~{\rm GeV}\, (''\diamond'' ), \, 325  ~{\rm GeV}\,
(''+'') ,\, 1000 ~{\rm GeV} \, (''.'')$, for  $\tan\beta=30$ (right panel).}
\label{fig2}
\end{figure}

In  Fig.~2, those portions of the parameter space which belong to different values of Q
are presented by curves ($'' + ''$,\,  $'' \times''$,\, $''.''$,\, $''\diamond''$ ), and
those curves correspond to different values of $\mu$.
A comparative look at ${\rm Q}=325 {\rm GeV}$  $(''+'')$, and  ${\rm Q}=600  ~{\rm GeV}$ $(''\times'')$
scales, in the low $\tan\beta$ regime (left panel), shows that
when ${\rm  Q}=325 {\rm GeV}$  $(''+'')$
and $\tan\beta=4$ (left panel), there are very few curves,
since the lower bound of  $\mu$ (corresponding to ${\rm  Q}=325 {\rm GeV}$)
starts from  a rather higher value of  $\mu$
($\mu=800{\rm GeV}$),
and, it is not possible to find any allowed region
below this value ($\mu\simlt 800{\rm GeV}$),
since the parameter space is constrained by the existing LEP bound on the lightest
Higgs boson mass \cite{LEP}.
Moreover, even at this value of $\mu$, one can not find
solutions in the full $\varphi_{kt}$ range, as will be shown in Fig.~3.
Therefore, the allowed range of the parameter space is quite restricted for this case.
At  ${\rm Q}=600 {\rm GeV}$ and  $\tan\beta=4$ (left panel),
$\rho_3$ decreases relatively as compared to the  former case.
However, as the allowed range of $\mu$ gradually gets widened, the number of curves increase
(for instance, the lower bound  of $\mu$  is  $750{\rm GeV}$ for this scale).
On the other hand, $\rho_3$ is the quite small at ${\rm  Q}=1000
{\rm GeV}$, and  $\tan\beta=4$ (left panel), as compared to the other two  scales (${\rm Q}=325 {\rm
GeV}$, ${\rm Q}=600 {\rm GeV}$). However, there is a gradual 
enlargement in the allowed range of $\mu$   
(The allowed range of $\mu$ for all values of Q will be discussed in detail,
in analyzing the $Q-\mu$ interdependence).  
The variation of  $\rho_3-\varphi_{kt}$ is similar 
in  the high  $\tan\beta$ regime (right
panel of Fig.~2). Considering the portion of the parameter space which corresponds to
${\rm  Q}=190 {\rm GeV}$ \, $(''\diamond'' )$, one notes that
the parameter space is quite widened, even  as compared to a larger
scale, ${\rm  Q}=325 {\rm GeV}$ \, $(''+'')$, in the low $\tan\beta$ regime.
On the other hand,  a comparision between
the left and  right panels, at ${\rm  Q}=325 {\rm GeV}$, also  shows that,
the number of curves, namely the allowed range of $\mu$  increases
gradually, as well as $\rho_3$, in the high $\tan\beta$ regime (right panel).

From the analysis of Figs.~1,~2,
we can conclude that the allowed range of $\mu$, as well as $\rho_3$
gradually increases with the increasing $\tan\beta$.
However, as the $\rho_3-\mu$ dependence is taken into consideration, 
it is seen that  some differences arise in connection with the different values
of Q.
Therefore, to understand the properties of different values of Q, and in particular 
their influences  on  $\rho_{3}$, as well as on $\mu$, 
we carry out the same analysis from a slightly different perspective in
Figs.~3,~4,~5  focusing on the behaviour of
the small, moderate and large values of Q,
under different values of $\mu$, in both regimes
\begin{figure}[htb]
\centering
\epsfxsize=6in
\epsffile{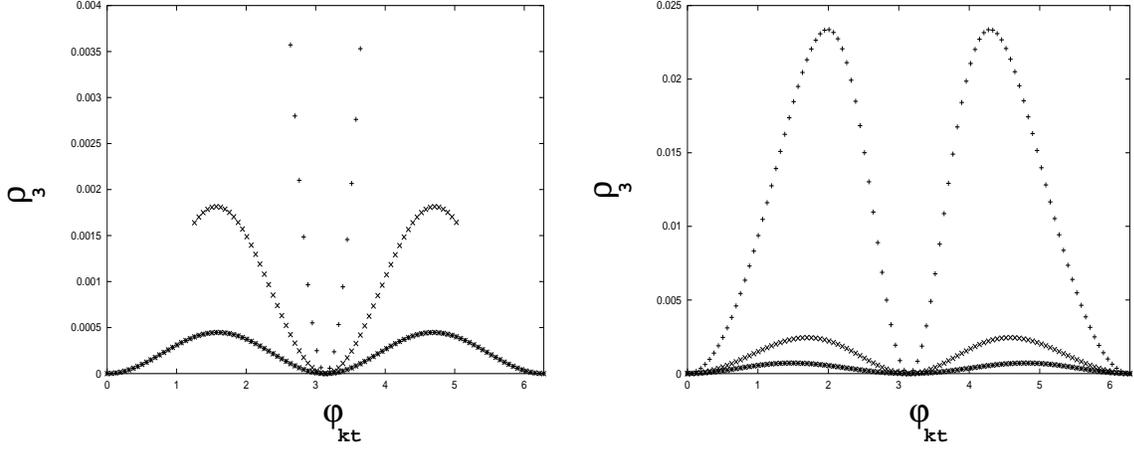}
\caption{The CP--odd composition ($\rho_{3}$) of the lightest Higgs boson as a
function of $\varphi_{kt}$, for  $\tan\beta=4$ (left panel), and $\tan\beta=30$ (right panel),
and $\mu=800 ~{\rm GeV}$,
where the top, middle and bottom curves correspond to ${\rm  Q} =325 ~{\rm
GeV}$ ($''+''$),  $600 ~{\rm GeV}$ ($''\times''$), $1000 ~{\rm GeV}$ ($''*''$) respectively.}
\label{fig3}
\end{figure}

Shown in Fig.~3,  is the variation of the CP--odd composition ($\rho_{3}$) of the lightest Higgs boson as a
function of $\varphi_{kt}$, for  $\tan\beta=4$ (left panel),
and $\tan\beta=30$ (right panel), when   $\mu=800 ~{\rm GeV}$.
The three different curves present three different values of Q  corresponding to
$ 325 ~{\rm GeV} \, ({\rm top \, curve} ), \, 600  ~{\rm GeV}\,({\rm middle
\, curve} ),
1000 ~{\rm GeV} \, ({\rm bottom \, curve} )$, respectively.
The same analysis has been carried out for $\mu=900 ~{\rm GeV}$ in Fig.~4,
and,  for $\mu=1000 ~{\rm GeV}$ in Fig.~5.
As is seen from  the left panel of  Fig.~3,
at  ${\rm  Q} =325 ~{\rm GeV}$, and  $ \mu=800 ~{\rm GeV}$ (the lower
bound of $\mu$ for this scale),
there are solutions only in a  small portion of parameter space
in the full $\varphi_{kt}$ range, which corresponds to   $[4 \pi/5, 6\pi/5]$ interval,
since  one can not find any allowed region satisfying the experimental
constraint \cite{LEP} beyond this interval.
At ${\rm  Q} =325 ~{\rm GeV}$, and $\mu=900 ~{\rm GeV}$ (left panel of Fig.~4), 
the allowed range of $\varphi_{kt}$
slightly widens. One notes that, for  ${\rm  Q} =325 ~{\rm GeV}$,  
the experimental constraint is satisfied in the full
$\varphi_{kt}$ range, only at  $\mu=1000 ~{\rm GeV}$ (left panel of Fig.~5).
On the other hand,  as the renormalization
scale increases,  for  instance, at ${\rm  Q} =600 ~{\rm GeV}$,
the allowed range of $\varphi_{kt}$ gradually gets widened, and
at  ${\rm  Q} =1000 ~{\rm GeV}$, there are solutions in the full
$\varphi_{kt}$ range. Similar observations can be made for  the
$\tan\beta=30$ case. However, as is seen  from the right panels of
Figs.~3,~4,~5 that the
experimental constraint is satisfied for all values of $\varphi_{kt}$ in  this regime.
\begin{figure}[ht]
\centering
\epsfxsize=6in
\epsffile{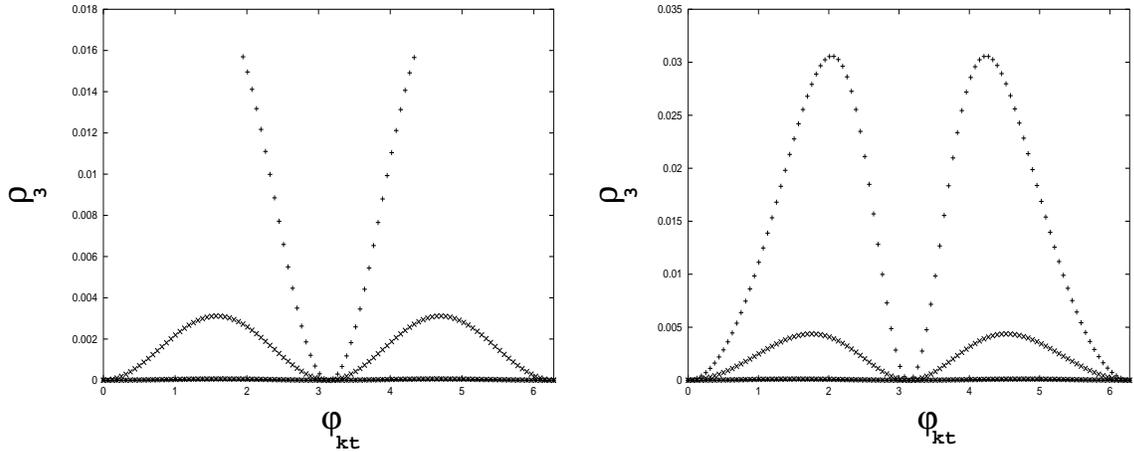}
\caption{The same as Fig.~3, but for  $\mu=900 ~{\rm GeV}$}
\label{fig4}
\vspace{0.25cm}
\end{figure}
\begin{figure}[ht]
\centering
\epsfxsize=6in
\epsffile{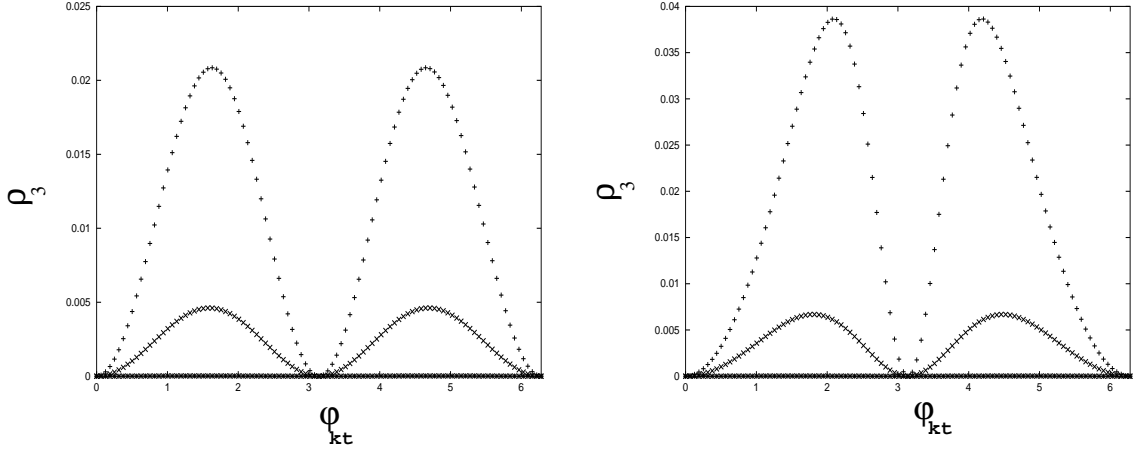}
\caption{The same as Fig.~3, but for  $\mu=1000 ~{\rm GeV}$}
\label{fig5}
\end{figure}

Moreover, as is shown in  the left panel of Fig.~4, for  ${\rm  Q} =325 ~{\rm GeV}$,
the maximal value of $\rho_3$
occurs at $\sim 0.004$ for $\mu=800 ~{\rm GeV}$. Then,
increasing slightly, and it reaches to $\sim 0.016$
at $\mu=900 ~{\rm GeV}$ (the left panel of Fig.~5). Finally, the  maximum
value of $\rho_3$ occurs at  $\sim 0.022$
at $\mu=1000 ~{\rm GeV}$ (left panel of Fig.~6 ).
On the other hand,  $\rho_3$ slightly decreases for ${\rm  Q} =600 ~{\rm GeV}$,
as compared to the  former  case (${\rm  Q} =325 ~{\rm GeV}$),
and  like the former case, its  $\rho_3$-$\mu$ dependence is of the same kind
(the larger   $\mu$, the  larger  $\rho_3$).
In passing to ${\rm  Q} =1000 ~{\rm GeV}$,
one notes from the left panel of Fig.~3 that,  $\rho_3$ decreases relatively
as compared to ${\rm  Q} =325 ~{\rm GeV}$, and ${\rm  Q} =600 ~{\rm GeV}$ cases,
and it  occurs maximally
at $\sim 0.0005$ at $\mu=800 ~{\rm GeV}$. However, unlike the  ${\rm  Q} =325 ~{\rm
GeV}$ and  ${\rm  Q} =600 ~{\rm GeV}$ cases,
$\rho_{3}$ (corresponding to ${\rm  Q} =1000 ~{\rm GeV}$)
is smaller  at  $\mu=900 ~{\rm GeV}$ than that of
$\mu=800 ~{\rm GeV}$, and
in fact, it  reaches to its minimal value at   $\mu=1000 ~{\rm GeV}$ (which
will be seen  in detail in analyzing the   $\rho_{3}-\mu$
interdependence of ${\rm  Q} =1000 ~{\rm GeV}$).

Therefore, the  comparative analysis of Figs.~3,~4 and 5 shows that,
there are differences in the $\rho_3$-$\mu$
dependence of Q in both low and high $\tan\beta$ regimes, in the sense that,
although  $\rho_3$  increases with increasing $\mu$
for  $Q= 325 ~{\rm GeV},\, \mbox{and},  \, Q=600  ~{\rm GeV}$, its dependence on $\mu$ changes
in passing to the higher renormalization scales
(for instance at $Q=1000 ~{\rm GeV}$,  $\rho_3$  decreases  with increasing
$\mu$). Therefore, one can deduce from the above analysis
that,  there should
be critical values of Q in both  low and high $\tan\beta$ regimes,
beyond which the $\rho_3$-$\mu$
dependence changes. To determine these portions of the parameter space,
it is necessary  to discuss the
variation of $\rho_3$ with $Q$, which will be carried out in the following.
\begin{figure}[htb]
\centering
\epsfxsize=6in
\epsffile{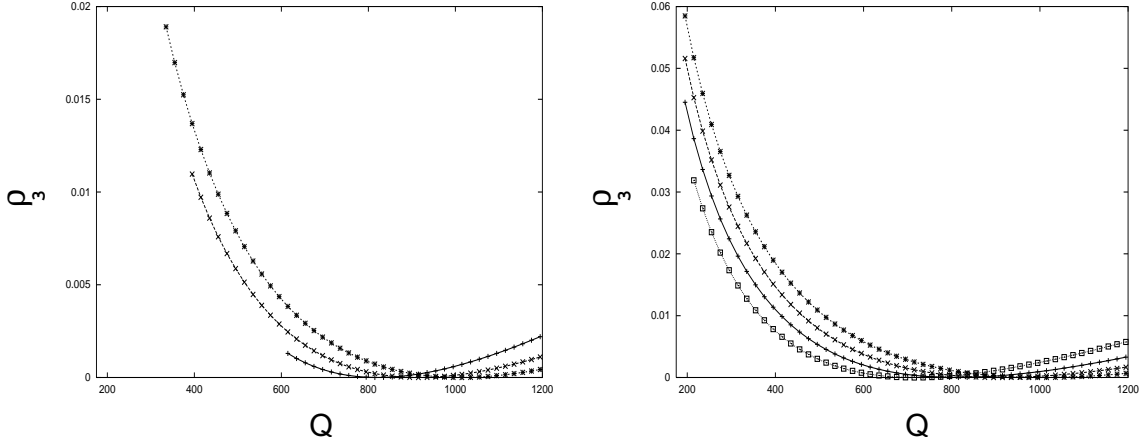}
\caption{The CP--odd composition ($\rho_{3}$) of the lightest Higgs boson
as a function of the renormalization scale ${\rm Q}$, at $\varphi_{kt}=3
\pi/2$, for  $\tan\beta=4$ (left panel), and  $\tan\beta=30$ (right panel),
with  different values of $\mu$  corresponding to 
$\mu=975\mbox{GeV}$\, $(''*'')$,  $875\mbox{GeV}$\, $(''\times'')$,
$775\mbox{GeV}$\, $(''+'')$, $675\mbox{GeV}$\, $(''\diamond'')$}
\label{fig6}
\end{figure}

Before analyzing the  dependence of Q  on $\rho_{3}$ in the full
$\varphi_{kt}$ range from a general point of view,
we focus on a smaller portion of a parameter space, for clearance.
Therefore, in Fig.~6, we choose a particular value of  $\varphi_{kt}$, namely,
we set $\varphi_{kt}= 3 \pi/2$. Moreover,
as Q changing from  $175 ~{\rm GeV}\, \mbox{to}\, 1200 ~{\rm GeV}$, we
let  $975\mbox{GeV}$\, $(''*'')$,  $875\mbox{GeV}$\, $(''\times'')$,
$775\mbox{GeV}$\, $(''+'')$, $675\mbox{GeV}$\, $(''\diamond'')$
values of $\mu$, for $\tan\beta=4$ (left panel), and  $\tan\beta=30$ (right  panel).
One notes that, there is no allowed region in the parameter space corresponding to
$\mu=675\mbox{GeV}$ for $\tan\beta=4$ (left panel). In fact,
it is not possible to find any
region in the parameter space below  $\mu \simlt 700$ for $\tan\beta=4$
(this bound is $\mu \simlt 500$ for  $\tan\beta=30$), due to the experimental  constraint on
the lower bound of the lightest Higgs mass \cite{LEP}. Thus,  in choosing the
different values of $\mu$, we try to form the most suitable combination.
In Fig.~6, for the ease of following we cut the vertical axes $\rho_{3}=0.02\%$
for $\tan\beta=4$ (left panel), and $\rho_{3}=0.06\%$ for $\tan\beta=30$
(right panel). However, as will be seen in Fig.~7,  it actually extends to  $0.025\%$
for $\tan\beta=4$, and  $0.18\%$ for $\tan\beta=30$  in the full
$\varphi_{kt}$ interval.
Although the analysis has been carried out for  a particular value of
$\varphi_{kt}$ ($\varphi_{kt}= 3 \pi/2$),
one notes that there are two portions in the parameter space.
In the first portion, $\rho_3$ decreases with decreasing $\mu$, then
after a certain value of Q, in the second portion, $\rho_3$  starts to increase
slightly with decreasing $\mu$. For instance, as is seen from the left panel of Fig.~6,
in the first portion of the parameter space ($Q\simlt 925\mbox{GeV}$),
the maximal value of $\rho_3$
occurs at  $\mu=975\mbox{GeV}$\, $(''*'')$,
and it decreases slightly with the decrease in $\mu$.
On the other hand, in the second portion of the parameter space
($Q \simgt 925\mbox{GeV} $), the maximal value of $\rho_3$ is smaller than that of the first region.
However,  $\rho_3$ is maximal  at $\mu=775\mbox{GeV}$\, $(''+'')$, and,
unlike the first portion of the parameter space, $\rho_3$ lessens as $\mu$
increases. For instance, $\rho_3$ is minimal at $\mu=975\mbox{GeV}$\,
$(''*'')$, in the
second portion. Similar observations can be made also for the high $\tan\beta$
regime. Clearly, with the analysis of Fig.~5,
we focus  on a particular value of $\varphi_{kt}$,
and  try to understand the $\rho_3$-$\mu$ dependence of Q, considering  different values
of $\mu$. In the following, we will carry out the
same analysis from a more general point of view, 
which will provide us to determine the critical values of Q more precisely,
and to understand the properties of the model
under concern in more detail. 
\begin{figure}[htb]
\centering
\epsfxsize=6in
\epsffile{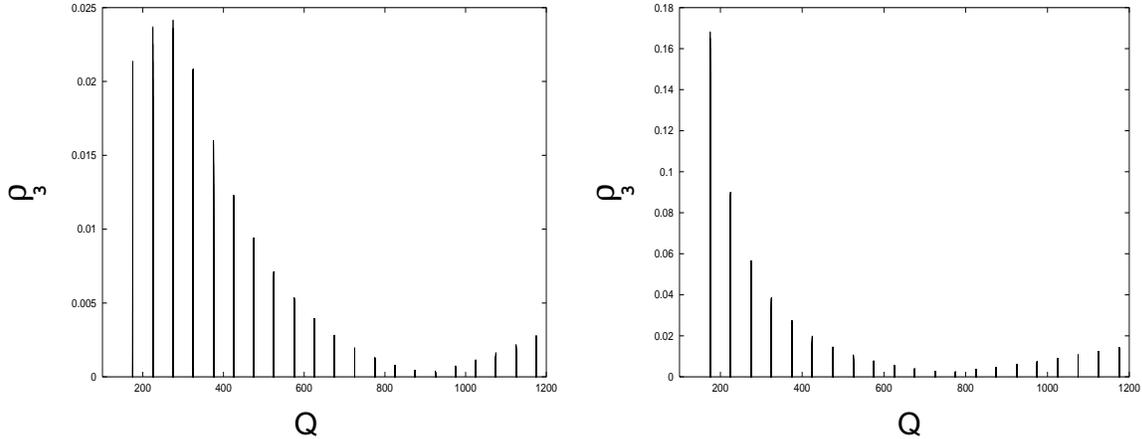}
\caption{The CP--odd composition ($\rho_{3}$) of the lightest Higgs boson
as a function of the renormalization scale ${\rm Q}$,
for  $\tan\beta=4$ (left panel), and  $\tan\beta=30$ (right panel).}
\label{fig7}
\end{figure}

In Fig.~7, we show the variation of  $\rho_{3}$ with
Q for  $\tan\beta=4$ (left panel), and  $\tan\beta=30$ (right panel)
in the full $\varphi_{kt}$ range.
In each panel, Q ranges  from $175 ~{\rm GeV} \,\, \mbox{to}\,\, 1200 ~{\rm
GeV}$, $\mu$  from $100~{\rm GeV}$ to $1000~\mbox{GeV}$.
Here, the vertical lines represent
different values of $\varphi_{kt}$, as well as different values of $\mu$, as
was indicated in Fig.~6.
As is seen from the left panel of Fig.~7, the maximum value of $\rho_{3}$
($\sim 0.025\%$) occurs at  $Q=275~{\rm GeV}$.
For larger values of ${\rm Q}$,  $\rho_3$ decreases gradually
until   $Q=925 {\rm GeV}$, as the supersymmetric spectrum decouples.
One notes that, beyond this point $\rho_3$ starts to increase slightly.
For smaller values of ${\rm Q}$, however,
the parameter space is constrained by the existing LEP bound on the lightest
Higgs mass \cite{LEP}. That is, 
$\rho_3$ gets smaller until $Q = 175~{\rm GeV}$, and
it is not possible to find any region in the parameter space
below this value ($ Q \simlt 175~{\rm GeV}$), for $\tan\beta=4$,
since this region is completely disallowed by the experimental bound
\cite{LEP}.
On the other hand, as is shown in the  right panel of Fig.~7,  the maximum value of
$\rho_{3}$
($ \sim 0.17\%$) occurs at  $Q = 175~{\rm GeV}$ for $\tan\beta=30$.
Similar to the low $\tan\beta$ regime, $\rho_3$ decreases with increasing $Q$
until $Q=775{\rm GeV}$, and beyond this point there is a small increase in
$\rho_3$ in the admitted range of Q. 

Following the analysis of  Figs.~3-7,
we would like to emphasize  that
in the first portion of the parameter space ($175\simlt Q \simlt 925$ for $\tan\beta=4$,
and $175\simlt Q \simlt 775$ for
$\tan\beta=30$), as  $\mu$ increases with increasing  $\rho_3$ (Figs.~3-6), 
$\rho_3$ decreases with increasing $Q$ (Figs.~6-7).
On the other hand, in the second portion of the parameter space
($925\simlt Q \simlt 1200$ for $\tan\beta=4$, and $775\simlt Q \simlt 1200$ for $\tan\beta=30$),
$\rho_3$  gradually increases  with  increasing  Q (Figs.~6-7),
and since this increase in  $\rho_3$ is compensated by the decrease in $\mu$,
unlike the first region, $\mu$ decreases with increasing $\rho_3$ (see Fig.~3-6). 
Therefore, the interdependence of $\mu-\rho_3$ changes in this region (see Fig.~6). 
However, the maximal values of  $\rho_3$
never exceeds $\sim 0.005 \%$ for $\tan\beta=4$, and   $\sim 0.02\%$ for
$\tan\beta=30$, in the second portion of  the parameter space.
Obviously if the lightest Higgs boson were carrying large enough CP-odd
composition, this would bring about new opportunities for observing the 
lightest Higgs in the near future. However, one notes that  
as the  CP violation effects are induced only radiatively 
in  the MSSM and in the model at hand, it is clear that any would-be observation 
of a  large  CP violating composition of the lightest Higgs boson at LHC, or NLC, or TESLA
excludes the MSSM in general, and the model under concern in particular.
\begin{figure}[htb]
\centering
\epsfxsize=6in
\epsffile{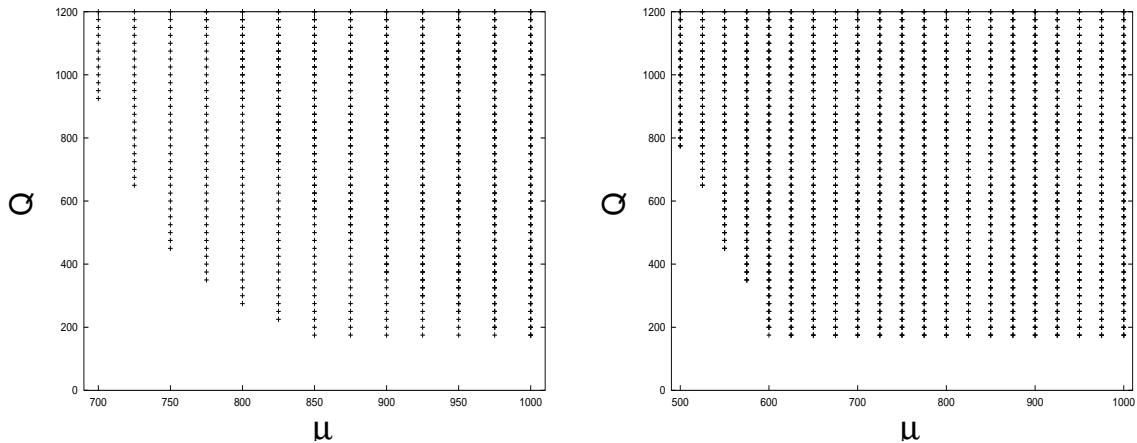}
\caption{The dependence of the renormalization scale Q
on $|\mu|$  for $\tan\beta=4$ (left panel), and $\tan\beta=30$ (right panel).}
\label{fig8}
\end{figure}

In Fig.~8, we show the interdependence of Q on $|\mu|$
for  $\tan\beta=4$ (left panel), and  $\tan\beta=30$ (right panel).
In each panel Q varies from $175 ~{\rm GeV} \,\, \mbox{to}\,\, 1200 ~{\rm
GeV}$, and  $\mu$  from $100~{\rm GeV}$ to $1000~\mbox{GeV}$.
As is seen from the left panel, in the first portion of the parameter space,
starting from $Q = 175 {\rm GeV}$,  the range of Q gets widened
up to $Q = 925 {\rm GeV}$, whereas
the lower allowed  bound of $\mu$  changes
from  $850 ~{\rm GeV} \,\, \mbox{to}\,\, 700~{\rm GeV}$
in this interval.
For instance, at $Q=325 ~{\rm GeV}$ and $\tan\beta=4$, the lower
bound of $\mu$ starts from $800~{\rm GeV}$ (see Fig.~3),
and,  at $600 ~{\rm GeV}$, it decreases to $750~{\rm GeV}$
in the same regime, as will be shown in Fig.~9.
On the other hand, in the second portion of the parameter space
($925 \simlt Q \simlt 1200$),
the allowed range of $\mu$
remains constant. That is,
all values of $\mu$ changing from
$700 ~{\rm GeV} \,\, \mbox{to}\,\, 1000 ~{\rm GeV}$ are allowed
for  $Q\simgt 925$ (for instance, the lower bound of $\mu$,
corresponding to   $Q = 1000 {\rm GeV}$ starts from
$700 ~{\rm GeV}$, as will be indicated in Fig.~10).
Q-$\mu$ interdependence is similar
at large $\tan\beta$ (right panel).
However, in this regime, as the range of Q gradually widens in the  $175 \simlt Q\simlt
775~{\rm GeV}$ interval,
the lower bound on $\mu$ changes
from  $600 ~{\rm GeV} \,\, \mbox{to}\,\, 500~{\rm GeV}$.
Above this interval ($Q\simgt 775$), all values of $\mu$, from  $500 ~{\rm GeV} \,\, \mbox{to}\,\, 1000 ~{\rm GeV}$
are allowed. For instance, the lower bound of $\mu$, corresponding to   $600
~{\rm GeV}$  starts from $\mu=550 ~{\rm GeV}$ at $\tan\beta=30$, whereas it is
$\mu=500 ~{\rm GeV}$, for  $Q = 1000 {\rm GeV}$ in the same regime, beyond which all values of
$\mu$ are allowed (see Figs.~9-10).     

We would like to remind  that  in the present analysis, the allowed range of the parameter space
is obtained by imposing the recent LEP constraint on the lightest Higgs mass ($m_{h_{3}}$)
which requires
$m_{h_{3}}\simgt 115\mbox{GeV}$ \cite{LEP}.
As is seen from both panels of Fig.~8,
the lower bound on $\mu$ starts from  $700\mbox{GeV}$ for $\tan\beta=4$ (left
panel), whereas this bound decreases to $500\mbox{GeV}$ for $\tan\beta=30$ (right
panel), and it is not possible to find any region in the
parameter space below these values ($\mu\simlt 700\mbox{GeV}$ for $\tan\beta=4$,
and $\mu \simlt500\mbox{GeV}$ for
$\tan\beta=30$),  since these regions of the parameter space are completely
disallowed  by the existing LEP bound on the lightest Higgs mass \cite{LEP}.
Moreover, it is worthwhile of mentioning that  
the present analysis includes the dominant top-stop
contributions,  as well as the bottom-sbottom effects,
whereas the two-loop corrections have  not been   taken into account. In this case,
the approximation used
here for the calculation still contains a theoretical error of several
$\mbox{GeV}$ \cite{brignole}. Therefore, the allowed range might change
mildly, in case of a higher precisional calculation.

With the former analysis, we have studied the properties of the model under concern,
as the Q-dependence is taken into the consideration. At this point,
we come back to the question mentioned in the
purpose of this work (whether one can find an appropriate limit of
reasonable agreement with the scale independent results \cite{boz}).
Therefore, in Fig.~9, we focus on two particular values of Q, namely
$Q=600~{\rm GeV}$, and  $Q=1000~{\rm GeV}$, and investigate their $\rho_3-\mu$
dependence in  both high and low $\tan\beta$ regimes.
\begin{figure}[htb]
\centering
\epsfxsize=6in
\epsffile{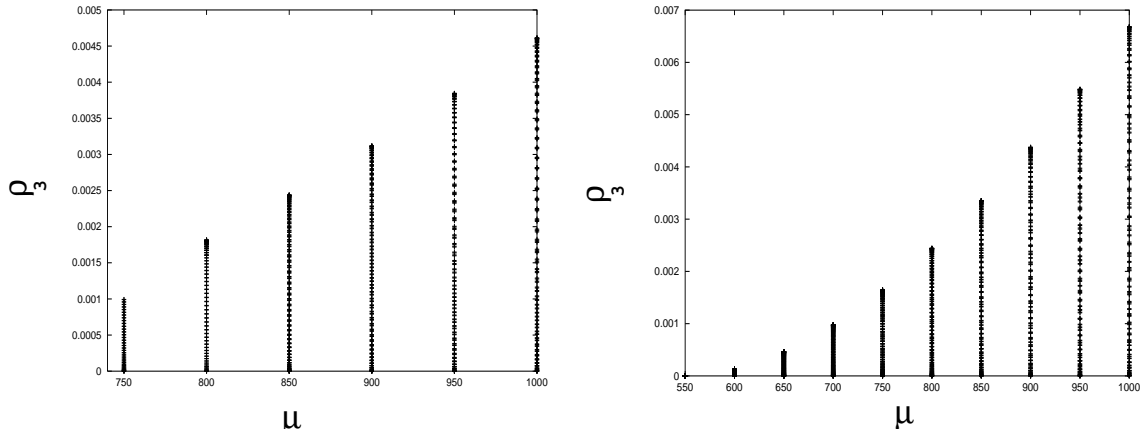}
\caption{The CP--odd composition ($\rho_{3}$) of the lightest Higgs boson as
a function of  $|\mu|$ for  $Q=600 ~{\rm GeV}$,  when  $\tan\beta=4$ (left panel),
 and  $\tan\beta=30$ (right panel) }
\label{fig9}
\end{figure}
\begin{figure}[htb]
\centering
\epsfxsize=6in
\epsffile{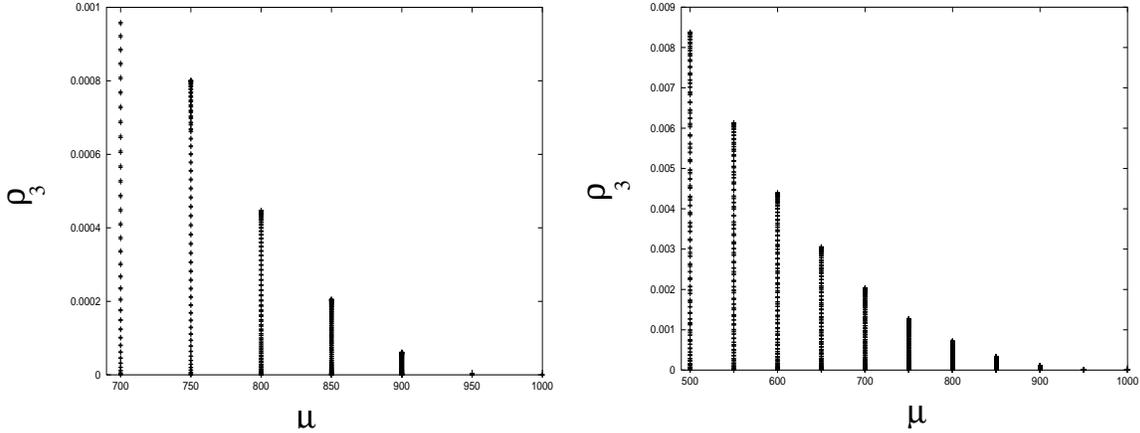}
\caption{The same as Fig.~9 , but  for  $Q=1000 ~{\rm GeV}$. }
\label{fig10}
\end{figure}

Depicted  in Fig.~9 is  the $|\mu|$ dependence of
$\rho_3$ for  $\tan\beta=4$ (left panel), and $\tan\beta=30$ (right panel)
at $Q=600~{\rm GeV}$. In Fig.~10, the same dependence is shown for
a larger scale, namely for $Q=1000~{\rm GeV}$.
In both panels, the vertical lines correspond to different phases
in the full $\varphi_{kt}$ range (see  Figs.~3,~4,~5).
A comparative glance at both panels shows that, in both high and
low  $\tan\beta$ regimes, the  $\rho_3-\mu$
dependence at  $Q=600  ~{\rm GeV}$ scale differs from that of   $Q=1000~{\rm GeV}$.
In fact, the   $\rho_3-\varphi_{kt}$ dependence corresponding to these scales has been analyzed in Figs.~3,~4,~5,
and  the critical values of Q (beyond which $\rho_3-\mu$ behaviour
changes) has been determined in Figs.~6,~7.
Therefore, such kind of behaviour is expected,
remembering that in the first  portion of the parameter space 
($175\simlt Q \simlt 925$ for $\tan\beta=4$, and $175\simlt Q \simlt 775$ for
$\tan\beta=30$),
$\mu$ increases with increasing  $\rho_3$ (Fig.~7), whereas it decreases with  increasing $\rho_3$
in the second portion of the total Q range ($925\simlt Q \simlt 1200$
for $\tan\beta=4$, and  $775\simlt Q \simlt 1200$ for $\tan\beta=30$).
As is seen from the left panel of  Fig.~ 9,
at  $Q=600~{\rm GeV}$, and   $\tan\beta=4$, the maximum value of $\rho_{3}$
($\sim 0.005\%$) occurs at  $\mu=  1000~{\rm GeV}$,
and for smaller  values of $\mu$, $\rho_3$ decreases
gradually. In passing to  $\tan\beta=30$ case (right panel),
one observes that  as $\rho_{3}$ increases slightly,
the allowed range of $\mu$ widens (the lower bound on $\mu$
decreases to  $\mu=550\rm{GeV}$).
On the other hand, as is seen from the left panel of Fig.~10, at  $Q=1000~{\rm GeV}$, and  $\tan\beta=4$,
the maximal value of
$\rho_3$ ($\sim 0.001\%$) occurs at  $\mu= 700~{\rm GeV}$, and
unlike the   $Q=600  ~{\rm GeV}$ case, for larger values of $\mu$,  $\rho_3$ decreases  gradually.
The variation of $\rho_{3}$ with $\mu$, for $\tan\beta=30$ 
(right panel of Fig.~10) is similar to what we have observed in the $\tan\beta=4$ 
regime (left panel of Fig.~10). However,  there is a gradual increase in $\rho_3$,
as well as the allowed range of $\mu$, for  $\tan\beta=30$.

We would like to note that the variation of $\rho_{3}$
with $\mu$ (Fig.~ 10), which corresponds to $Q=1000~ {\rm
GeV}$, is very similar to what we have found in \cite{boz}. This particular result
shows that the scale dependence of the radiative corrections is sufficiently
suppressed for the renormalization scale  $Q=1000~ {\rm GeV}$.
On the other hand, consideration of various renormalization scales, all
being around the weak scale, lead us to a wealth of CP violation opportunities.
This is especially suggested by  Fig.~9, where the
percentage  CP-odd composition of the lightest Higgs gradually increases
with $\mu$.
\begin{figure}[htb]
\centering
\epsfxsize=6in
\epsffile{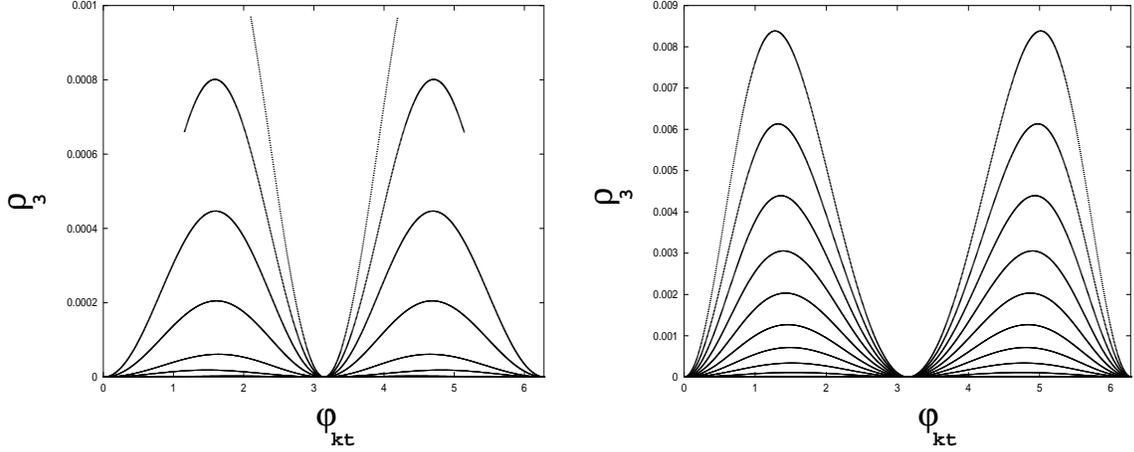}
\caption{The dependence of the CP--odd composition ($\rho_3$)
of the lightest Higgs boson on  $\varphi_{kt}$, corresponding to
$Q=1000 \mbox{GeV}$,  for $\tan\beta=4$ (left panel), and  $\tan\beta=30$ (right panel)}
\label{fig11}
\end{figure}

As mentioned above, among the values of Q changing from
$175 ~{\rm GeV} \,\mbox{to}\, 1200 ~{\rm GeV}$,
of particular interest  is the $Q=1000 {\rm GeV}$ where the comparision
between  the results of this work and that of \cite{boz} is possible,
and a reasonable agreement with \cite{boz} can be found.
Shown in Fig. 11, is the $\varphi_{kt}$ dependence of
the percentage CP--odd  compositions of the lightest
Higgs boson ($\rho_{3}$), corresponding to   $Q=1000 ~{\rm GeV}$, for
$\tan\beta=4$ (left panel), and $\tan\beta=30$ (right  panel).
As both panels of the figure suggests
the maximal value of $\rho_3$
occurs at  $\sim  0.001\%$ for $\tan\beta=4$ (left panel), and increasing slightly
it reaches to $\sim  0.009\%$ for $\tan\beta=30$ (right  panel),
in the full $\varphi_{kt}$ range. As has been shown in Figs. 3,4,5
each curve in the figure  represents different values of $\mu$.
For instance, in the $\tan\beta=4$ regime (left panel)
the curve on the top corresponds to  the lower bound of $\mu$
($\mu=700\mbox{GeV}$), for which $\rho_3$ reaches its maximal (see Fig.~10), and similarly
from the top curve to the bottom with the increase in  $\mu$, $\rho_{3}$
decreases. Similar to observations made for the left panel, one can discuss
the high $\tan\beta$ regime (right panel), where  the top curve corresponds
to the lower bound of $\mu$ ($\mu=500\mbox{GeV}$), for
$\tan\beta=30$ (see Fig.~10).
One notes that, the  $\rho_3$ reaches maximally to  $\sim  0.009\%$ 
in the full $\varphi_{kt}$ range, for all values of  $\tan\beta$
changing from 4 to 30.

As is explained in the Introduction,
considering only the dominant top-stop quark loops, and being the results of
Q-independent, the variation of $\rho_3$
with  $\varphi_{kt}$ is analyzed in \cite{boz}, and
it has been  found that the maximal value of
$\rho_3$  never exceeds $0.0013\%$ for all values of $\tan\beta$ changing 4 to 30 in the
full $\varphi_{kt}$ range 
(the maximal values of   $\rho_3\sim 0.0003\%$  for $\tan\beta=4$, and $\rho_3\sim
0.0013\%$  for $\tan\beta=30$ in \cite{boz}) 
The dependence of  $\rho_{3}$ on  $\varphi_{kt}$
(see Fig. 10), which corresponds to  $Q=1000 ~{\rm GeV}$,
is very similar to that of \cite{boz}. 
However, one notes that  maximal value of $\rho_3$ slightly increases
in this work, as compared to the former analysis.

Among the particles contributing to the
one-loop radiative corrections, the dominant ones come from the top quark
and top squark loops, provided that $\tan\beta \simlt 50$,
(for which case the bottom Yukawa coupling is quite small to
give significant contributions). The Yukawa interactions due to scalar
bottom quarks can be significant only for very large $\tan\beta$
values. Therefore, by the inclusion of the bottom-sbottom loops $\rho_3$ increases
slightly  as expected, however,
it never exceeds  $0.009\%$,
for all values of $\tan\beta$ changing from 4 to 30.
\begin{figure}[htb]
\centering
\epsfxsize=6in
\epsffile{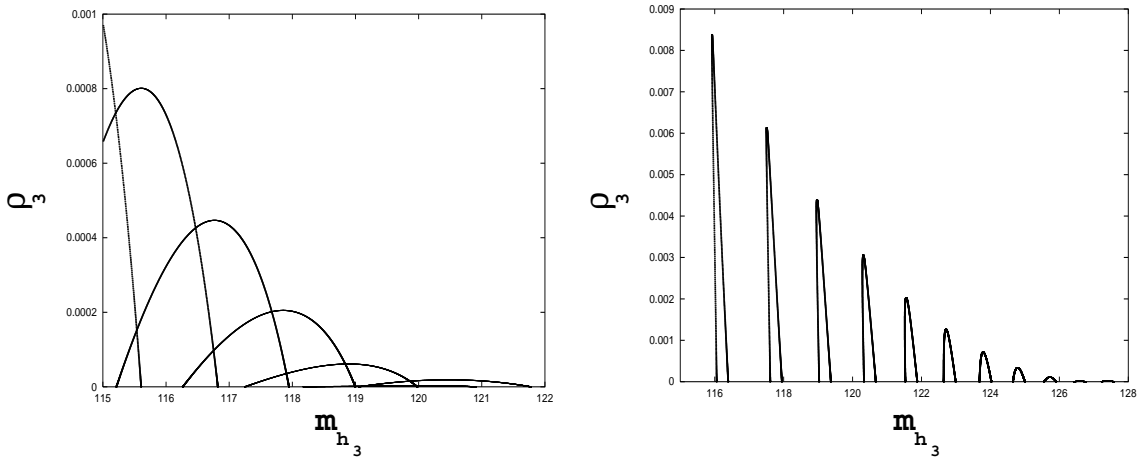}
\caption{The dependence of the  mass of  the lightest Higgs boson
($m_{{h}_{3}}$)  on  $\varphi_{kt}$, corresponding to
$Q=1000 \mbox{GeV}$,  for $\tan\beta=4$(left panel), and  $\tan\beta=30$
(right panel)}
\label{fig12}
\end{figure}

Finally, depicted in Fig.~12, is  the variation of $\rho_3$ with
$m_{h_{3}}$, corresponding to  $Q=1000 ~{\rm GeV}$, for
$\tan\beta=4$ (left panel),
and  for $\tan\beta=30$ (right panel).
The  decreasing curves in each panel represent different values of
$\varphi_{kt}$, and $\mu$ as well. The maximal value of
$\rho_3$ occurs at  $\sim  0.001\%$, and correspondingly
at $\mu=700\mbox{GeV}$ (see Figs.~10,~11),
Then, it decreases rapidly with the increasing mass, and
decreasing $\mu$. For instance,
it reaches to  $\rho_3 \sim  0.0005\%$ at  $\mu=800\mbox{GeV}$ (see
Figs.~3,~10,~11), for $\tan\beta=4$ (left panel).
Similar observations can be made for the right panel, in which case
the maximal value of  $\rho_3 \sim  0.009\%$ occurs at $\mu=500\mbox{GeV}$
(see Fig.~11), and like the former case,  it decreases relatively with the 
increasing mass again. For instance,  $\rho_3 \sim  0.005\%$, at  
$\mu=800\mbox{GeV}$ (see Figs.~3,~10,~11). Finally, 
it reaches far below  $\rho_3 \sim  0.001\%$ for $m_{{h_{3}}}=127 \mbox{GeV}$.

As both panels of the figure shows 
that lighter the Higgs boson ($h_3$)
larger its CP--odd composition. 
Therefore, as the  $\rho_3$-$m_{h_{3}}$ dependence
(which corresponds to  $Q=1000 ~{\rm GeV}$)
suggests,
any possible increase in the lower experimental bound 
of the lightest Higgs mass will imply reduced CP--odd composition,
being in a reasonable agreement with the results of \cite{boz}.

Before concluding, we would like to note that
we have concentrated on a particular scale $Q=1000~ {\rm GeV}$,
for which case the results are very similar to what we have found
in the scale-independent case \cite{boz}.
In fact, in both low and high  $\tan\beta$ regimes,
$Q=1000~ {\rm GeV}$,
is an approximately intersectional value
of  the second portion of the
parameter space ( $925\simlt Q \simlt 1200$ for $\tan\beta=4$,
and  $775\simlt Q \simlt 1200$ for $\tan\beta=30$), beyond which  the $\rho_3-\mu$ dependence changes.
Moreover, if we consider the remaining interval in  the second portion of the parameter space
$1000\simlt Q\simlt 1200$, similar results can be obtained.
Naturally, in the $1000\simlt Q\simlt 1200$ interval,
 with the decrease in $\mu$, $\rho_3$ increases gradually,
which is a characteristic behaviour of the
second portion of the parameter space (see Fig.~7).
However $\rho_3$ never exceeds   $  \sim 0.02\%$ even at  $Q=1200~ {\rm GeV}$. 
Therefore, one  can conclude that
there is a strong influence of Q on the
$\rho_3$-$\mu$ interdependence in the first portion of the parameter
space ($175\simlt Q \simlt 925$ for $\tan\beta=4$, and  $175\simlt Q \simlt
775$ for $\tan\beta=30$). 
On the other hand, the scale dependence of the radiative corrections is
sufficiently suppressed in the second portion of the parameter space,
in particular, $1000 \simlt Q  \simlt 1200$,  
interval, for both low and high $\tan\beta$ regimes.

\section{Conclusion}
In this work, we have mainly concentrated on
the percentage CP odd-compositions of the
lightest Higgs boson, with the inclusion of the bottom-sbottom contributions,
for all values of Q ranging from top mass to ${\rm TeV}$
scale. We would like to briefly summarize the results:

\begin{enumerate}
\item
A comparative glance at both low and high $\tan\beta$
regimes shows that
as Q increases, the percentage  CP--odd composition of the lightest Higgs
($\rho_3$)
decreases until  $Q=925 {\rm GeV}$ for $\tan\beta=4$,
and $Q=775 {\rm GeV}$ for  $\tan\beta=30$.
Beyond these points, $\rho_3$ gradually starts to increase in the admitted 
range of Q. However, it never exceeds  $0.17\%$ for all values of  
Q changing from   $175 ~{\rm GeV} \,\, \mbox{to}\,\, 1200 ~{\rm GeV}$.
As compared to its percentage CP--even compositions
this value is so small to cause observable effects.
Therefore, lightest Higgs remains essentially CP even.

\item
The present analysis
differs from that of \cite{boz}, in the sense that the latter does not
depend on Q explicitely since the D-term as well as the bottom-sbottom
contributions are not taken into the consideration.
However, in the appropriate limit, namely for $Q=1000{\rm GeV}$,
a  comparision between  the results is possible,
and a reasonable agreement with \cite{boz} can be found.
Moreover, similar observations can be made 
for the second portion of the parameter space,
particularly in the  $1000 \simlt Q  \simlt 1200$ interval,
which is  an allowed intersectional region in  both low and high $\tan\beta$ 
regimes.
Therefore, one can conclude that there is a strong influence of Q on the
$\rho_3-\mu$ interdependence in the first portion of the parameter space.
On the other hand, the scale dependence of the radiative corrections is sufficiently
suppressed for the second portion of the parameter space,
particularly in the  $1000 \simlt Q  \simlt 1200$ interval.
However, consideration of 
various renormalization scales, all
being around the weak scale, lead us to a wealth of CP violation opportunities.

\item
The underlying model provides  a quite restricted parameter 
space due to the naturalness requirements, as well as a simultaneous
solution both to the  the strong CP and $\mu$--problems.  
\end{enumerate}

\noindent
M. B would like to thank the Scientific and Technical Research
Council of Turkey (T\"{U}B{\.I}TAK) for partial support under the project,
No:TBAG2002(100T108).

\end{document}